\documentclass[]{article}

\pdfoutput=1
\usepackage{authblk}
\usepackage{fullpage}
\usepackage{amsmath}
\usepackage{booktabs}

\usepackage{caption}
\usepackage{graphicx}

\usepackage[hidelinks]{hyperref}
\usepackage[round, semicolon]{natbib}
\usepackage{multirow}
\usepackage{lscape}

\title{Subsurface Imaging Dataset Acquired at the Garner Valley Downhole Array Site using a Dense Network of Three-Component Nodal Stations}
\author[1]{Joseph P. Vantassel}
\date{}
\author[2]{Jodie A. Crocker}
\author[3]{Brady R. Cox }
\author[4]{Khiem Tran}
\affil[1]{Virginia Polytechnic Institute and State University}
\affil[2]{The University of Texas at Austin}
\affil[3]{Utah State University}
\affil[4]{University of Florida}

\begin{document}

\maketitle

\begin{abstract}
There is a growing need to characterize the engineering material properties of the shallow subsurface in three-dimensions for advanced engineering analyses. However, imaging the near-surface in three-dimensions at spatial resolutions required for such purposes remains in its infancy and requires further study before it can be adopted into practice. To enable and accelerate research in this area, we present a large subsurface imaging dataset acquired using a dense network of three-component (3C) nodal stations acquired in 2019 at the Garner Valley Downhole Array (GVDA) site. Acquisition of this dataset involved the deployment of 196, 3C nodal stations positioned on a 14 by 14 grid with a 5-m spacing. The array was used to acquire active-source data generated by a vibroseis truck and an instrumented sledgehammer, and passive-wavefield data containing ambient noise. The active-source acquisition included 66 vibroseis and 209 instrumented sledgehammer source locations. Multiple source impacts were recorded at each source location to enable stacking of the recorded signals. The active-source recordings are provided in terms of both raw, uncorrected units of counts and corrected engineering units of meters per second. For each source impact, the force output from the vibroseis or instrumented sledgehammer was recorded and is provided in both raw counts and engineering units of kilonewtons. The passive-wavefield data includes 28 hours of ambient noise recorded over two night-time deployments. The dataset is shown to be useful for active-source and passive-wavefield three-dimensional imaging as well as other subsurface characterization techniques, which include horizontal-to-vertical spectral ratios (HVSRs), multichannel analysis of surface waves (MASW), and microtremor array measurements (MAM). 
\end{abstract}

\pagebreak

\section*{Introduction}

The ability to reliably and non-invasively measure engineering material properties of three-dimensional (3D) subsurface volumes at high spatial resolution has the potential to revolutionize site characterization. While work has been on-going in this area over the past several years (e.g., Fathi et al., \citeyear{fathi_three-dimensional_2016}; Tran et al., \citeyear{tran_3-d_2019}), progress towards a practical 3D imaging approach that can be applied routinely for engineering analyses remains out of reach. The authors believe that this is due in part to the lack of publically available, high-quality, large-scale, field datasets for near-surface, 3D imaging in locations of engineering interest. As evidence for this, new geophysical methods (e.g., Butzer et al., \citeyear{butzer_3d_2013}; Wang et al., \citeyear{wang_3d_2019}) are often applied only on synthetic data due to the lack of high-quality field datasets and invasive data for verification. In response, this work presents a high-quality, large-scale, field datase for near-surface, 3D imaging acquired at the Garner Valley Downhole Array (GVDA) site in southern California.

\subsection*{Site Overview}

The GVDA is located in southern California approximately 70 miles northeast of San Diego and 90 miles southeast of Los Angeles. The GVDA exists in a seismically active region 7 km from the San Jacinto Fault and 35 km from the San Andreas Fault. Instrumentation at the GVDA began in 1989 and includes multiple ground motion recording stations at the surface and at depth (i.e., a downhole/borehole array configuration) \citep{archuleta_garner_1992}. The site is particularly interesting from an earthquake engineering perspective, as quite a few previous studies \citep{steidl_what_1996, bonilla_borehole_2002, teague_measured_2018, afshari_california_2019, tao_insights_2019, vantassel_multi-reference-depth_2019} have attempted to match the frequency-dependent site amplifications observed at the GVDA using one-dimensional (1D) ground response analyses (GRAs), but with rather limited success. In fact, this has led several authors to conclude that the GVDA site amplification must be significantly influenced by non-1D behaviors. As such, this makes the GVDA site a particularly interesting test bed for 3D site characterization efforts. 

\begin{figure}[!]
    \centering
	\includegraphics[width=0.5\textwidth]{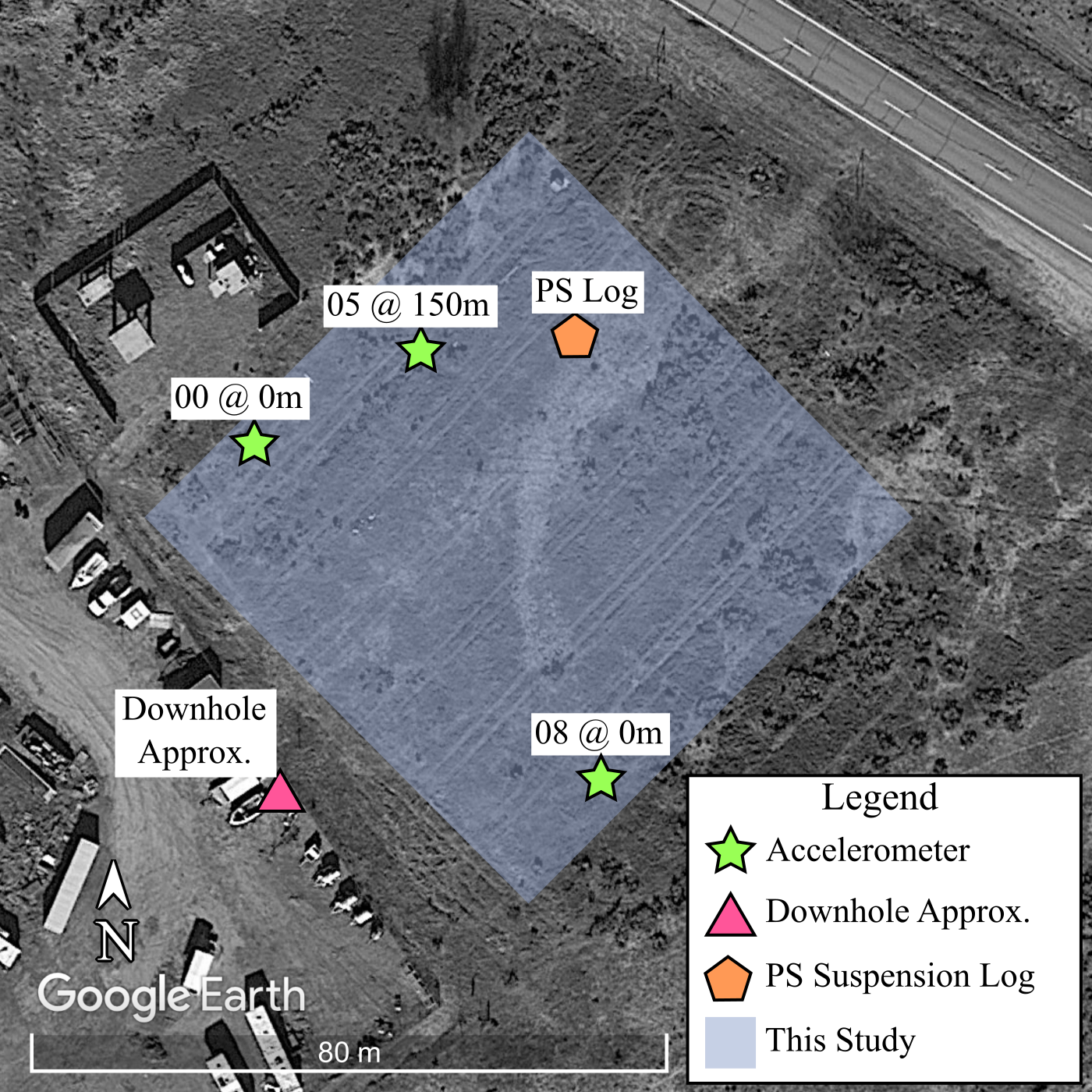}
	\caption{Plan view of the Garner Valley Downhole Array (GVDA) site with the locations of three of the permanent ground motion accelerometers indicated with stars, the approximate location of the seismic downhole test performed by Gibbs (\citeyear{gibbs_near-surface_1989}) indicated with a triangle, the location of the PS suspension logging performed by Steller (\citeyear{steller_new_1996}) indicated with a pentagon, and the shaded area indicating the region of interest for the present study.}
	\label{fig:1}
\end{figure}

A plan view of the GVDA site is shown in Figure \ref{fig:1}. Stars denote three of the site's permanent ground motion accelerometers; 00 and 08 at the surface (i.e., 0 m depth), and 05 located at 150 m depth. The GVDA site has been the focus of a number of previous invasive site characterization efforts. Of those efforts, two were performed to sufficient depth to be of interest for GRA-type analyses: the PS suspension logging performed by Steller (\citeyear{steller_new_1996}) and the downhole test performed by Gibbs (\citeyear{gibbs_near-surface_1989}). The location of the PS suspension logging is known fairly accurately (within a few meters) and is shown with a pentagon in Figure \ref{fig:1}. The location of the downhole test, however, is not well known. Its approximate position was estimated from a map provided in the report by Gibbs (\citeyear{gibbs_near-surface_1989}) and is shown in Figure \ref{fig:1} with a triangle, although the true location may be anywhere within a 30 m radius of the location indicated.

Several previous non-invasive site characterization efforts have also been conducted at the GVDA site. These include spectral analysis of surface waves (SASW) tests performed by Stokoe et al. (\citeyear{stokoe_sasw_2004}), a 3D full waveform inversion (FWI) dataset collected with 1C geophones reported by Fathi et al. (\citeyear{fathi_three-dimensional_2016}), and multi-channel analysis of surface waves (MASW) and microtremor array measurement (MAM) tests documented by Teague et al. (\citeyear{teague_measured_2018}). Of these efforts, only the data reported by Fathi et al. has been openly published \citep{bielak_t-rex_2012}, and while it is an early and extremely valuable dataset for active-source FWI studies, it is significantly less expansive than the present dataset in terms of the number of receivers, receiver components, number and type of active-sources, and inclusion of passive-wavefield data.  As the array and receiver locations used for these studies are too numerous to show in Figure \ref{fig:1}, we refer the reader to the publications noted above for detailed descriptions of the experimental layouts relative to the GVDA site's instrumentation.

\subsection*{Site Geology}

The geology of the GVDA site includes three primary geologic layers: alluvial sediments (AL), decomposed granite (DG), and competent granite (GR) \citep{archuleta_garner_1992}. The AL contains a mixture of silty sand, clayey sand, and silty gravel \citep{archuleta_garner_1992, youd_geotechnical_2004}. Lake Hemet, the result of the Hemet Dam whose shoreline is located approximately 500 m to the west of the downhole array \citep{hill_geology_1981}, serves to maintain a saturated soil depth between 1 and 3 m below the ground surface depending on the season \citep{steidl_what_1996}, with some having measured it as deep as 5 m \citep{gibbs_near-surface_1989}. The bottom depth of the AL has been estimated using a variety of geotechnical and geophysical techniques. These tests include: downhole testing by Gibbs (1989), which estimated the bottom of the AL at approximately 18 m; standard penetration testing (SPT) performed by Pecker and Mohammadioun (\citeyear{pecker_downhole_1991}), which also estimated the bottom of AL at about 18 m; PS suspension logging performed by Steller (\citeyear{steller_new_1996}), which estimated a gradual transition between the AL and DG between 10 and 20 m; and spectral analysis of surface waves testing performed by Stokoe et al. (\citeyear{stokoe_sasw_2004}), which estimated the transition at approximately 24 m. Additionally, cone penetration tests (CPTs) performed by Youd et al. (\citeyear{youd_geotechnical_2004}) estimated the bottom of the AL deposit to range from 18 to 25 m across the site. Most recently, multichannel analysis of surface waves (MASW) and microtremor array measurements (MAM) by Teague et al. (\citeyear{teague_measured_2018}) estimated the thickness of the AL between 15 and 20 m. The range of estimates of the AL's thickness seems to imply that the interface between the AL and the DG at GVDA may vary significantly across the site, hinting at non-1D site conditions. Underlying the AL is the DG. The estimated bottom depth of the DG, much like the AL deposits, varies between sources. Gibbs (1989) estimated the depth to GR as approximately 65 m; Archuleta et al. (\citeyear{archuleta_garner_1992}), who reinterpreted the data from Gibbs (\citeyear{gibbs_near-surface_1989}), estimated it between 40 and 45 m; Steller (\citeyear{steller_new_1996}) estimated it at approximately 87 m; and Teague et al. (\citeyear{teague_measured_2018}) estimated it between 40 and 70 m. The potential for high variability in AL and DG layers at the site prompted the acquisition of the high-quality, large-scale imaging dataset whose presentation and documentation is the primary focus of this work.

\subsection*{Overview of the Dataset}

The dataset documented in this work was collected at the GVDA site between 6 – 10 October 2019. It involved the deployment of 196, 3-component (3C) nodal seismic stations borrowed from IRIS PASSCAL. The nodes were organized on a 14 x 14 grid at a 5-m spacing, producing an array of 65 m x 65 m. As shown in Figure \ref{fig:1}, the array encompassed the spatial location of the PS suspension logging by Stellar (\citeyear{steller_new_1996}) and the three aforementioned permanent seismic accelerometers, and was as close as possible to the approximate location of the downhole testing by Gibbs (\citeyear{gibbs_near-surface_1989}). The nodal array was used to record both active-source and passive-wavefield data. The former includes 209 source locations with an instrumented sledgehammer and 66 source locations with a vibroseis, whereas the latter includes 28 hours of ambient noise. All of the data discussed in this work has been archived and made publically available through DesignSafe-CI \citep{rathje_designsafe_2017} under the project titled, ``Active-Source and Passive-Wavefield Nodal Station Measurements at the Garner Valley Downhole Array" \citep{vantassel_active-source_2023}.

\section*{Instrumentation}

\begin{figure}[!]
    \centering
	\includegraphics[width=0.8\textwidth]{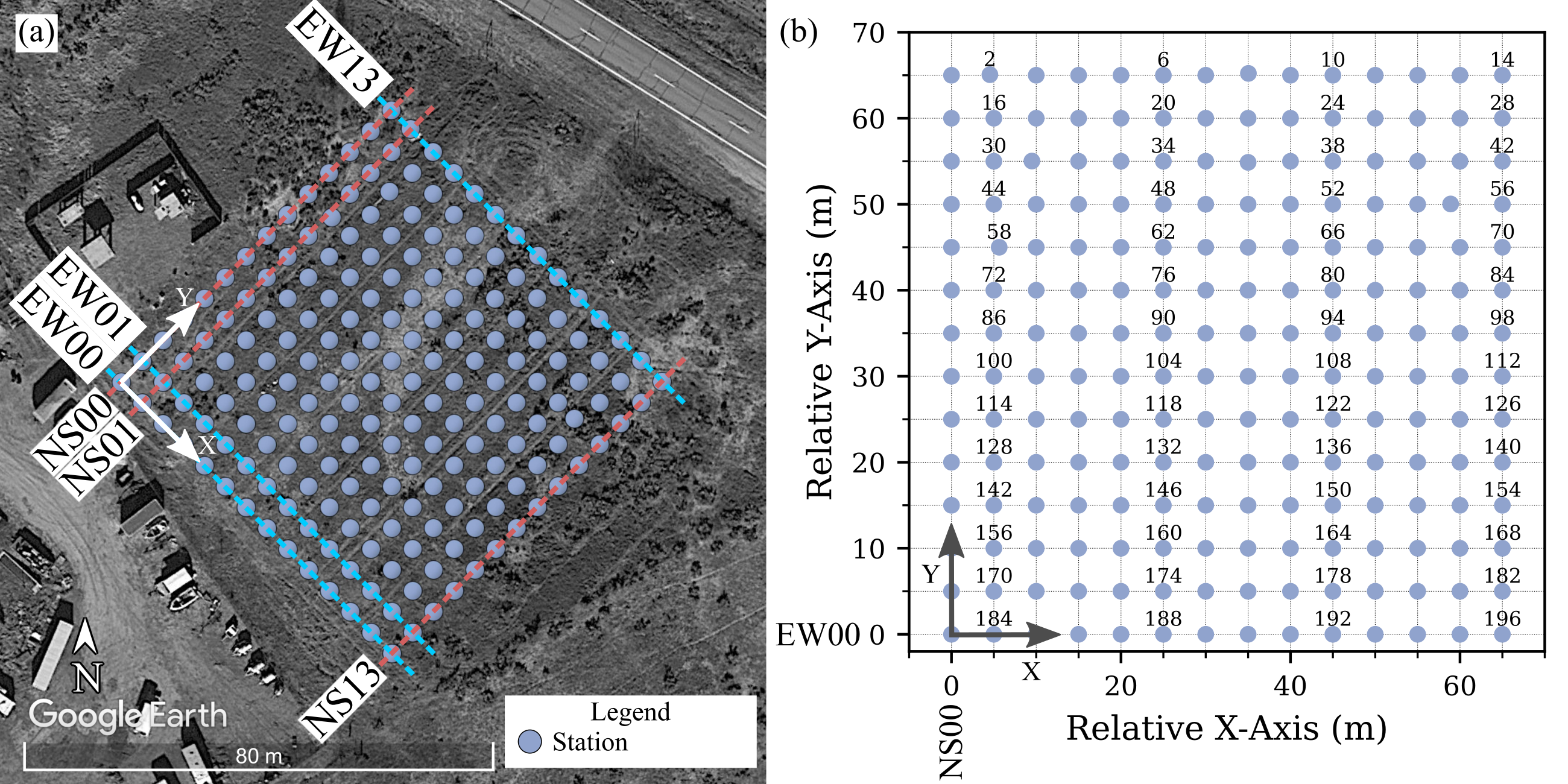}
	\caption{Panel (a) shows the as-built location of all 196, 3-component nodal stations deployed at the Garner Valley Downhole Array (GVDA) site. Each line of the square array is denoted with a four-character designation, with the first two characters denoting its orientation (NS or EW) and the second two its position from the array's western-most corner (00 to 13). Panel (b) shows the as-built array on its relative coordinate system with the sequential array position numbers of select stations indicated directly above their corresponding circle symbol. The orientation of the relative coordinate system shown in panel (b) is indicated with x-y arrows in panel (a).}
	\label{fig:2}
\end{figure}

\begin{figure}[!]
    \centering
	\includegraphics[width=0.5\textwidth]{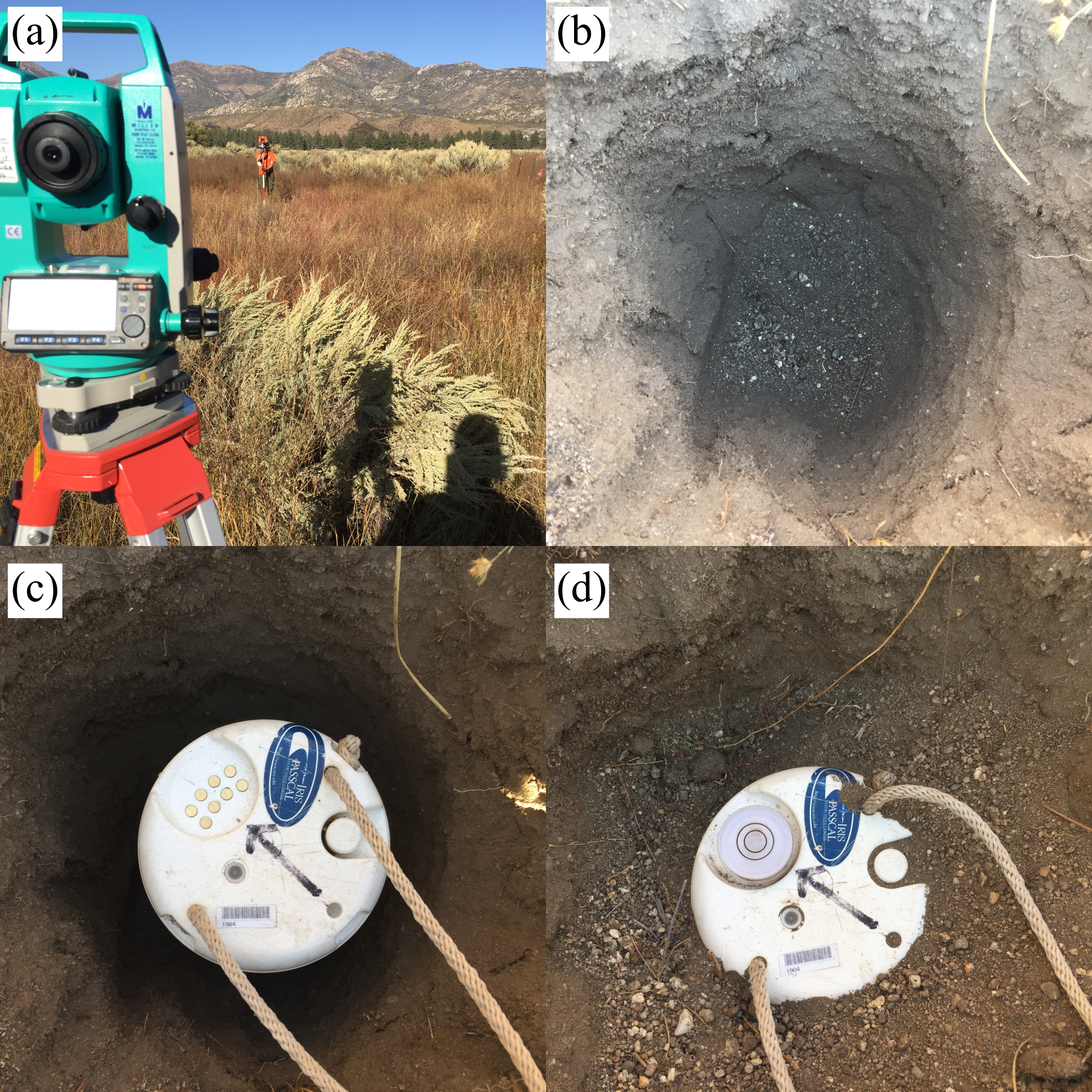}
	\caption{The 3-component nodal station installation process involved: (a) surveying each position to within a 5 cm accuracy using a total station, (b) excavating a hole for each station approximately 18 cm deep and 15 cm in diameter, (c) placing the sensor in the hole and orienting it to magnetic north, and (d) backfilling and leveling the senor.}
	\label{fig:3}
\end{figure}

The 196 nodal stations were arranged at the site in a 14 x 14 grid at a 5 m spacing, as shown in Figure \ref{fig:2}a. The square array was oriented approximately 45 degrees clockwise from north to allow it to include the ground motion accelerometers and PS suspension logging test location, and to avoid as much as possible on-the-ground obstructions. The first step of installing the array was to establish a relative coordinate system. As access to the site was through a gate near the array's western-most corner, it was decided this corner should serve as relative coordinate (0, 0), with positive x-values pointing to the southeast and positive y-values to the northeast. The x-y arrows in Figure \ref{fig:2}a denote the orientation of the array's relative coordinate system, with the layout of the array on that relative coordinate system shown in Figure \ref{fig:2}b. Each line of the square array is denoted with a four-character designation, with the first two characters denoting its orientation (NS or EW) and the second two its position from the array's western-most corner (00 to 13). In Figure \ref{fig:2}b, the sequential array position numbers of select nodal stations are indicated directly above their corresponding circle symbol. These sequential array position numbers are used to designate each station in the published dataset. To establish the location of all 196 stations in the field, the following procedure was used. First, a total station surveying instrument was setup at relative location (0, 0) and a bearing was selected to denote the array's relative y-axis. With the bearing set, all 13 locations along that line were surveyed into positon. The process was then repeated by rotating the total station 90 degrees to the east to set the bearing of the array's relative x-axis, and all 13 locations along that line were surveyed into position. The total station was then moved to relative coordinate (5 m, 0 m), orthogonality was ensured through pre-established fore- and back-sights, and the second line was surveyed into position. The process continued using this same procedure until all 196 stations were surveyed into position. Note that the surveying process was non-trivial due to the aforementioned obstructions and significant plant life at the site. A picture of the surveying process is shown in Figure \ref{fig:3}a. Note the knee-high grass and other larger plant life in the photo's background. It took approximately 10 hours to survey all 196 stations to an accuracy of 5 cm, with surveying starting on the afternoon of 6 October 2019 and ending late in the afternoon of 7 October 2019. All dates and times are listed in Pacific Daylight Time (PDT) unless noted otherwise. The as-built locations of all stations, including those that needed to be moved to avoid obstructions, are provided as part of the dataset in both relative (i.e., x-y) and absolute coordinates (i.e., latitude-longitude). Figures \ref{fig:2}a and \ref{fig:2}b show the as-built locations of all stations.

Following the layout of the array, holes were excavated by hand to bury the nodal stations. The holes were excavated approximately 18 cm deep and 15 cm in diameter. A picture of a typical hole is shown in Figure \ref{fig:3}b. An interesting point to note is that the near-surface material was highly variable. Some areas of GVDA were similar to a clean sand and easy to excavate with a posthole digger, whereas others were heavily cemented and required the use of a breaker bar. The process of excavating the holes took a crew of three people approximately 8 hours to complete. Excavation of the holes began on the afternoon of 7 October 2019 and concluded at about noon on 8 October 2019. Once the excavations were complete, the 3C nodal stations were placed into the excavated holes, oriented to magnetic north (see Figure \ref{fig:3}c), leveled, and backfilled (see Figure \ref{fig:3}d).  Because the nodal stations were buried, each nodal station's large central spike was removed to accelerate installation. Following installation, the node activation process involved manually connecting to each node, linking its serial number with a preplanned deployment layout, booting up the station, and confirming it achieved a GPS timing lock and was operating correctly. The process of installing and activating the stations took a crew of five approximately five and a half hours to complete (three and a half hours to install and two hours to activate). All stations were activated by 18:00 PDT on 8 October 2019.

The nodal stations deployed at GVDA were Fairfield Nodal ZLand 3C equipped with three orthogonal 5-Hz geophones (two horizontal and one vertical), an internal battery, and a GPS time-synced digitizer with an accuracy of +/- 10 microseconds. The nodal stations were borrowed from the PASSCAL Instrumentation Center as part of the project, ``Data Acquisition for 2020 Blind Trial of Surface-based Site Characterization Methods" led by Alan Yong \citep{yong_data_2019}. For their deployment at GVDA, the stations were configured with a passive mode setup that included a 24 dB gain and 2000 Hz sampling rate (0.5 ms time step). The stations were left to record continuously in passive mode through the duration of the experiment, including during active-source data acquisition. The decision to leave the stations in a passive mode configuration, rather than switching them between two configurations (one for active with a low or zero gain and one for passive with a higher gain), was done as a matter of convenience in the field, as each switch between modes would have required at least two hours. Leaving the stations in the passive mode configuration allowed more time for active-source data acquisition, while, as discussed in detail later, only moderately complicating post-acquisition processing. In hindsight, the authors believe trading post-processing complexity for faster field acquisition was worthwhile, as it allowed the acquisition of more data in the limited time available. Importantly, while preparing the data for publication, the authors have seen no evidence that the choice to use a pure passive-mode deployment had any negative effect on the final data quality.

\section*{Active-Source Data Acquisition}

\begin{figure}[!]
    \centering
	\includegraphics[width=0.8\textwidth]{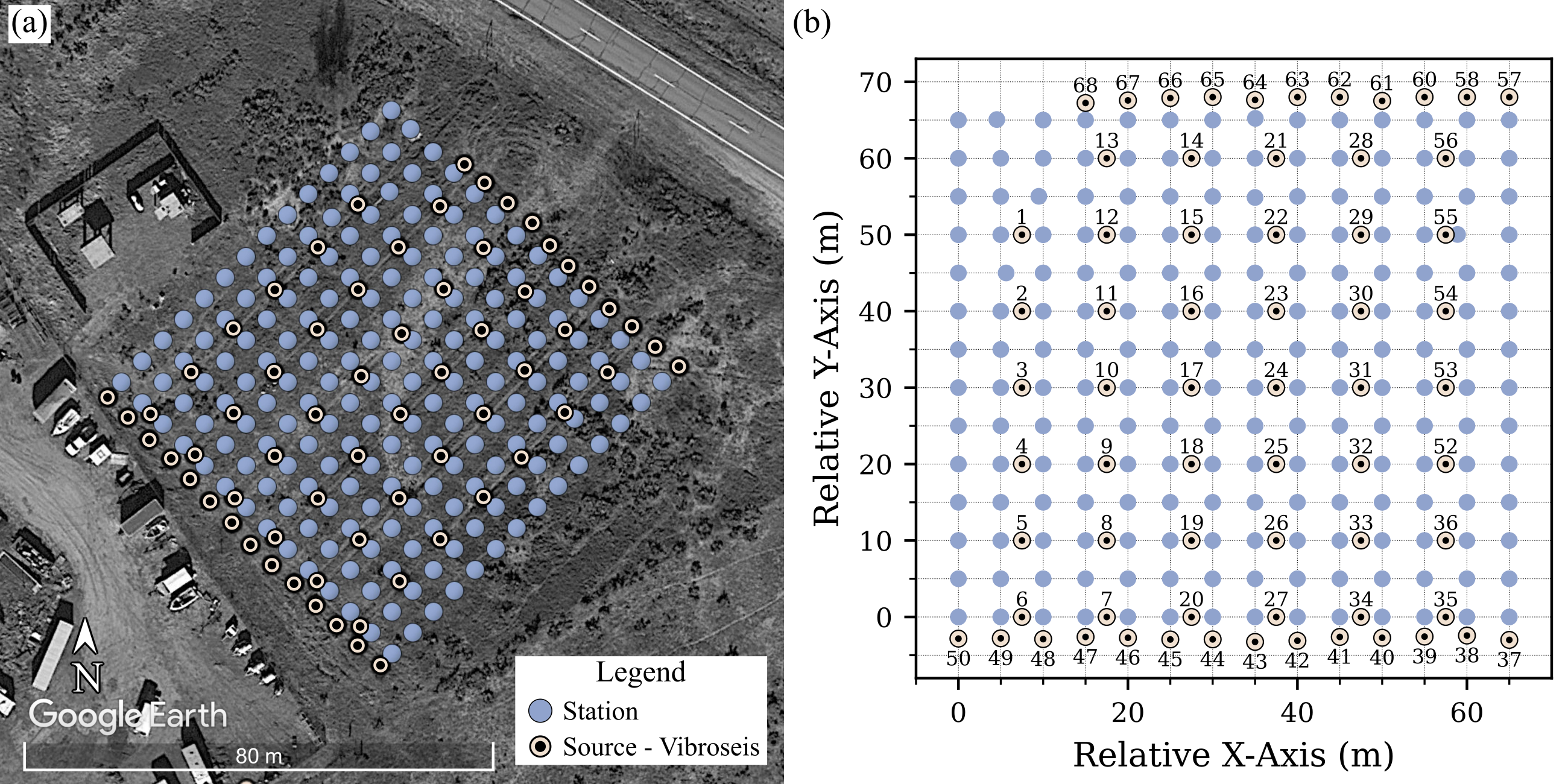}
	\caption{Panel (a) shows a plan view of the GVDA site with the 66 (41 interior; 25 exterior) vibroseis source locations relative to the 196 nodal stations. Note that it was not possible to operate the vibroseis along the array's northwestern and southeastern edges due to the presence of fencing and construction debris. Some other regularly spaced source locations also had to be skipped due to access restrictions. Panel (b) shows the vibroseis source locations with their numeric identifiers relative to the nodal stations on the site's relative coordinate system. Note that the vibroseis source numbers are not sequential, with some numbers missing. Missing numbers correspond to other source locations used in the field but not related to the dataset presented here.}
	\label{fig:4}
\end{figure}

Active-source data acquisition began at 09:00 PDT on 9 October 2019 with the vibroseis source. The vibroseis source used in this study was the highly mobile shaker truck Thumper from the NHERI@UTexas experimental facility \citep{stokoe_nheriutexas_2020}. Thumper was used to shake in the vertical direction with a linear frequency sweep from 5 to 30 Hz over a 12 second duration. Thumper was run in high-force-output mode, which at the GVDA site was able to reliably produce approximately 30 kilonewtons (kN) of ground force. During shaking, a Nanometrics' Centaur digitizer was connected directly to Thumper's electronics (i.e., into vibeout) to record the force output at a sampling rate of 2000 Hz (0.5 ms time step). Using a digitizer like the Nanometrics' Centaur that is GPS time synchronized was imperative to ensure a reliable point of reference between the source and the nodal stations, as no time break or seismic event recorder was used in the field to ``trigger" data acquisition on the stations. Note that extracting records for individual source events, therefore, had to be done as part of post-acquisition processing and will be discussed in detail later. As shown in Figure \ref{fig:4}a, Thumper was used to shake at 41 locations inside the array (i.e., interior source locations) and 25 locations outside of the array (i.e., exterior source locations) (66 in total). Coordinates for all Thumper shaking locations are tabulated in the published dataset. Note that it was not possible to test along the array's northwestern and southeastern edges and at some locations to the north due to the presence of fencing and construction debris. Figure \ref{fig:4}b shows the vibroseis source locations with their numeric identifier relative to the nodal stations on the site's relative coordinate system. Note that the vibroseis source numbers are not sequential, with some numbers missing. Missing numbers correspond to other source locations used in the field but not related to the dataset presented here. At each source location three identical shakes were performed to allow subsequent stacking of the measured signals. Active-source data acquisition with Thumper took approximately six hours to complete, concluding at approximately 15:00 PDT on 9 October 2019.

\begin{figure}[!]
    \centering
	\includegraphics[width=0.8\textwidth]{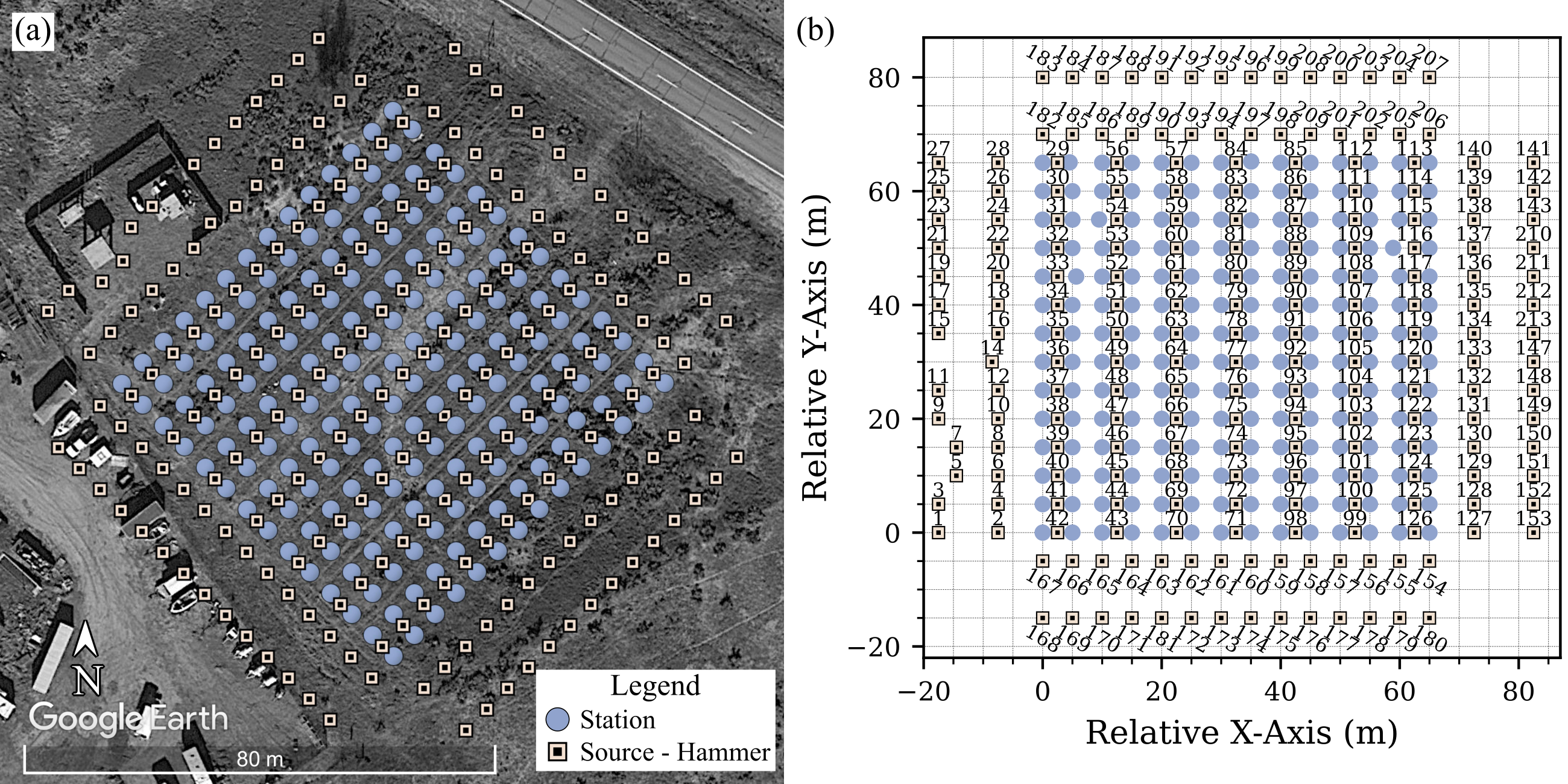}
	\caption{Panel (a) shows a plan view of the GVDA site with the 209 (98 interior; 111 exterior) hammer source locations relative to the 196 nodal stations. Note that the source locations on the northwestern side of the array had to be adjusted from their planned locations and, in one case, cancelled altogether to avoid equipment associated with the GVDA. Panel (b) shows the hammer source locations with their numeric identifiers relative to the nodal stations on the site's relative coordinate system. Note that the hammer source numbers are not sequential, with some numbers missing. Missing numbers correspond to other source locations used in the field but not related to the dataset presented here.}
	\label{fig:5}
\end{figure}

Following the vibroseis data acquisition, the sledgehammer source was used. The sledgehammer used in this study was a 5.4 kg instrumented sledgehammer from PCB Piezotronics. At each hammer source location, the hammer was swung vertically against an aluminum strike plate. As with the vibroseis source, a Nanometrics' Centaur digitizer was used to record the signal from the instrumented sledgehammer's load cell with a sampling rate of 2000 Hz (0.5 ms time step).  The force output and frequency content from the instrumented sledgehammer varied with the operator and the ground conditions, but on average at the GVDA site, the sledgehammer produced a peak force output of approximately 20 kN. As shown in Figure \ref{fig:5}a, the sledgehammer source was used at 98 interior source locations and at 111 exterior shot locations (209 in total). Note that the source locations on the western side of the array had to be adjusted from their planned locations and in one case cancelled altogether to avoid equipment associated with the GVDA. Figure \ref{fig:5}b shows the hammer source locations with their numeric identifier relative to the nodal stations on the site's relative coordinate system. As with the vibroseis source numbers, the hammer source numbers are not sequential, with some numbers missing. Missing numbers correspond to other source locations used in the field but not related to the dataset presented here. Coordinates for all of the hammer source locations are tabulated in the published dataset. At each location at least five source impacts were performed to allow subsequent stacking of the measured signals. The entire process took approximately 6 hours to complete (from 15:00 to 19:00 PDT on 9 October 2019 and from 08:00 to 10:00 PDT on 10 October 2019).

\section*{Post-Acquisition Processing}

At the conclusion of data acquisition, the stations were deactivated, excavated, cleaned, and packaged for shipment back to IRIS PASSCAL. The entire pickup and cleaning process took approximately 5 hours. Once the stations were received by IRIS PASSCAL the data was downloaded, archived on the IRIS Data Management Center (DMC), and provided to the project team. As all of the data for the GVDA experiment were recorded in a passive mode configuration, there were two main post-acquisition processing tasks: (1) downloading, post-processing, and organizing the passive-wavefield data in an easy-to-reuse format, and (2) identifying when each active-source event occurred using the source recordings, extracting the corresponding receiver data from the IRIS DMC, and providing the extracted events in an easy-to-reuse format.

The first and easier of the two tasks was to download and post-process the passive-wavefield data. There were two passive-wavefield recording periods. The first was a 15-hour block from 18:00 PDT on 8 October 2019 to 09:00 PDT on 9 October 2019 and the second was a 13-hour block from 19:00 PDT on 9 October 2019 to 08:00 PDT on 10 October 2019. Following standard practice, the passive wavefield data in our dataset is archived in coordinated universal time (UTC), not PDT, and therefore the aforementioned blocks correspond to 01:00 UTC to 16:00 UTC on 9 October 2019 and 02:00 UTC to 15:00 UTC on 10 October 2019, respectively. For archiving the passive-wavefield data on DesignSafe, two formats were under consideration: comma-separated value (CSV) and miniSEED (a subset of the Standard for the Exchange of Earthquake Data (SEED) format). Ultimately, it was decided that the widely used miniSEED format would allow for the greatest ease-of-reuse, as most ambient-noise-focused geophysical software accepts miniSEED as input, and there also exists several open-source libraries to read miniSEED (e.g., \emph{obspy}, \emph{libmseed}) for writing the data into other formats. In addition, using a binary-based format like miniSEED (rather than a text-based format like CSV) minimizes the disk space required to hold the data. To enable users to easily access subsets of the data, one miniSEED file was saved per station (i.e., all three components in a single file) per hour. Two post-acquisition processing steps were performed on the raw passive-wavefield records prior to writing the results to miniSEED. The first step was to clarify the miniSEED channel codes provided in the miniSEED metadata. The channel codes provided from the IRIS DMC were GP1, GP2, and GPZ, where the band code ``G" denotes a sensor sampled above or equal to 1000 Hz, but less than 5000 Hz, and with a corner frequency greater than 0.1 Hz; the instrument code ``P" denotes a very short period seismometer with a natural frequency greater than or equal to 5 Hz; and the orientation codes ``1", ``2", and ``Z" refer to an orthogonal horizontal component at a non-traditional orientation, a second orthogonal horizontal component at a non-traditional orientation, and the vertical component, respectively. Because all passive-wavefield data was downsampled from 2000 Hz (0.5 ms) to 200 Hz (5 ms) during the download from the IRIS DMC and stations in the field were oriented to magnetic north, the channel codes were updated from GP1, GP2, and GPZ to EPE, EPN, and EPZ, respectively. Note the band code ``E" denotes a sensor sampled above or equal to 80 Hz but less than 250 Hz and the orientation codes ``E" and ``N" denote the east and north horizontal components, respectively. The second post-processing step was to merge and resample the passive-wavefield recordings to ensure a single time series (i.e., trace) per component per hour, as components can be split into different traces during acquisition due to a station reacquiring GPS lock. When a station reacquires GPS lock it is able to correct for any drift that has occurred in its internal clock since its last GPS lock. While this correction ensures that the timings provided by the digitizer are as accurate as possible, it also requires a new trace with a new corrected start time to be initiated. As a result, a three-component station that has recently regained a GPS lock will contain six traces rather than the expected three (i.e., three from before the GPS lock was reacquired and three after the GPS lock was reacquired). The purpose of the post-processing step discussed here is to combine the split traces and resample the merged data to ensure the correct number of samples are provided in each hour-long trace. In practice, this involved using the open-source Python package \emph{obspy} \citep{the_obspy_development_team_obspy_2022} to read the raw miniSEED file into a Stream object, calling the merge method with the argument $fill\_value=``interpolate"$, calling the interpolate method with the arguments $method=``lanczos"$ and $a=20$, and writing the merged and resampled data to miniSEED. The download and post-processing of the passive-wavefield data took approximately 25 hours running on 5 threads via a script launched from the DesignSafe-CI's JuptyerHub. Note that more threads were not used to avoid overloading the IRIS DMC's servers. The passive-wavefield dataset contains 5488 miniSEED files (1 file per hour per station * 28 hours * 196 stations) with a total size of 93 GB. Note that no instrument corrections have been applied to the passive-wavefield data and, as such, the data stored in the miniSEED files is in the raw unit of counts. For those interested in applying instrument corrections to the passive wavefield data to recover engineering units (e.g., particle velocity in meters per second (m/s)), the information required to do so is provided as part of the dataset and the correction procedure is described later in the context of the active-source data.

\begin{figure}[!]
    \centering
	\includegraphics[width=1.0\textwidth]{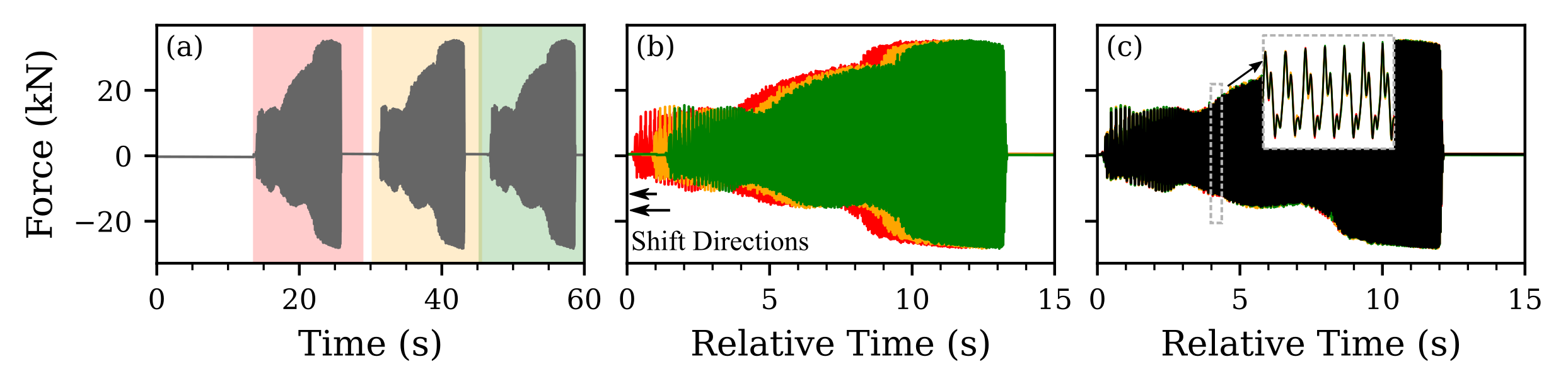}
	\caption{Example of the process used to align individual vibroseis shots prior to stacking. Panel (a) shows the signal recorded from the vibroseis for sixty seconds following the start time noted in the field at vibroseis source location 57. The three shaded areas denote the manually-selected, 15-second windows that contain the three vibroseis sweeps performed at this location. Panel (b) shows the three, 15-second windows plotted relative to their selected start time from panel (a). The direction of the cross-correlation time shifts relative to the first recorded shot are indicated with black horizontal arrows. Panel (c) shows the three source signals aligned using the time shifts determined from cross-correlation. The solid black line indicates the average or ``stack" of the three source signatures.}
	\label{fig:6}
\end{figure}

\begin{figure}[!]
    \centering
	\includegraphics[width=1.0\textwidth]{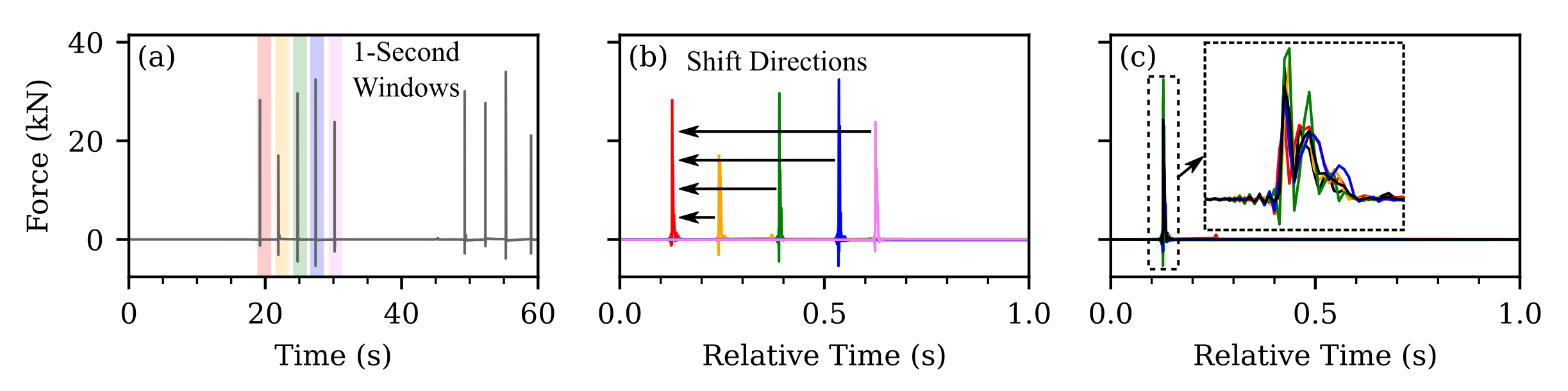}
	\caption{Example of the process used to align individual sledgehammer shots prior to stacking. Panel (a) shows the signal recorded from the instrumented sledgehammer for sixty seconds following the start time noted in the field at hammer source location 168. The five shaded areas denote the manually-selected, 1-second windows that contain the five source impacts performed at the source location. Panel (b) shows the five, 1-second windows plotted relative to their selected start time from panel (a). The direction of the cross-correlation time shifts relative to the first recorded shot are indicated with black horizontal arrows.  Panel (c) shows the five source signals aligned using the time shifts determined from cross-correlation. The solid black line indicates the average or ``stack" of the five source signatures.}
	\label{fig:7}
\end{figure}

The second, and more challenging of the two post-processing tasks, was to identify when each active-source occurred using the source recordings from the vibroseis and hammer, download and extract the corresponding receiver data, and organize the resulting active-source data in an easy-to-reuse format. During acquisition, handwritten field notes were kept for each active-source location that included the approximate start time (to the nearest minute), the number of shots performed, and which, if any, of the shots were thought to be of low quality. These notes were transcribed to an electronic format and reviewed twice to ensure their accurate reproduction. The times were then used as a reference point to preview the recorded source signatures for any given source location. Examples of what typical source signatures looked like are shown in Figures \ref{fig:6}a and \ref{fig:7}a for the vibroseis and hammer, respectively. A time of zero seconds denotes the start of the minute noted on the field datasheet. The amplitudes of each source have been converted from units of counts, as recorded on the digitizer, to engineering units of kN. The process of converting the source signature data to engineering units required a two-step conversion. First, the data was converted from counts to millivolts (mV) using the conversion factor from the Centaur digitizer of 2.50E-3 mV per count. Second, the voltage data was converted from mV to kN using the conversion factor from Thumper of 2.14E-2 kN per mV or from the instrumented sledgehammer of 4.17E-3 kN per mV. In Figure \ref{fig:6}a, three clear vibroseis chirps are shown in the 60-second time block, and in Figure \ref{fig:7}a, five clear hammer source impacts are visible between approximately 15 and 30 seconds (note that another cluster of shots corresponding to the next source location are visible in the same 60-second time block starting at approximately 47 seconds).

The individual times of each source impact per source location needed to be identified. To do this, we manually selected the approximate start time of each source impact, shown as the left-hand border of the 15-second and 1-second shaded windows in Figures \ref{fig:6}a and \ref{fig:7}a, respectively. Note that while this process could likely have been fully automated, the choice of doing it manually was made for two main reasons. The first was that several of the source locations, especially the hammer source locations, included extra shots (i.e., more than five) that were not listed on the datasheet (a consequence of miscounting in the field), and as such required human-level interpretation to separate the shots. The second was to ensure each source signature was visually reviewed for quality assurance. The process to approximately select the beginning time for all 1.3k source signatures took roughly 6 hours. These approximate, manually selected beginning times for each shot became the new zero times for each shot on a relative time scale (refer to Figures \ref{fig:6}b and \ref{fig:7}b). Then, they were cross-correlated to identify the number of time shifts required to align them for subsequent stacking. For those unfamiliar with cross-correlation, it involves repeatedly computing the element-wise product of two zero-padded time series where one of time series is systematically shifted past the other. The number of shifts that results in the largest product indicates the number of time shifts required to align the two time series. This process is shown in Figures \ref{fig:6}b and \ref{fig:7}b, where all source signatures are plotted relative to their picked times from Figures \ref{fig:6}a and \ref{fig:7}a, respectively, and arrows denote the direction and magnitude of their shifts calculated using cross-correlation relative to the first source signature. These shifts were then applied to align the source signatures prior to stacking. The aligned source signatures and the stacked source signature are shown in Figures \ref{fig:6}c and \ref{fig:7}c. Note that the black line indicates the average or ``stacked" source signature and it overlaps so well with the individual aligned shots (particularly in regards to the vibroseis signals, which are more consistent with one another) that it is difficult to view them all. With the source signatures shifted in preparation for stacking, it was now possible to determine the absolute start time for each source impact by taking the nearest minute written in the field, adding to it the manually picked time relative to that minute, and finally adding the cross-correlation time shifts. The aforementioned process provided the absolute time that denoted the start of each source impact required for relating the source and receiver station data.

While the absolute times derived above could have been used to extract each source record directly from the IRIS DMC, an alternate but equivalent approach was used. This alternate approach involved downloading short blocks of time before and after each source start time noted in the field (one for each station for each source position). Having the raw data stored locally allowed revisions to the post-acquisition processing workflow without requiring the data to be downloaded again from the IRIS DMC (a time consuming process). To provide data for the most likely processing scenarios, it was decided to include the waveform recordings from all stations for all interior shots (e.g., for FWI) and to provide waveform recordings associated with its corresponding line (see Figure \ref{fig:2}a) for all exterior shots (e.g., for MASW). Note that while exterior shots can be used for FWI (e.g., to provide longer ray paths) and the interior shots for MASW (e.g., to provide better spatial sampling), the authors decided the selected approach represents a reasonable simplification that is consistent with most studies. To reduce the size of the active-source data, it was downsampled from 2000 Hz (0.5 ms) to 400 Hz (2.5 ms) during its download from the IRIS DMC. With the data downloaded, the waveforms recorded by each receiver for each source impact were extracted using the cross-correlated source time shifts described above and reviewed. The review process included two data-quality checks. The first was a check on the stacking procedure itself that involved stacking the signals using the closest receiver rather than the source signature. The results from this check were time domain stacks of the receiver signals that were found to be identical to those obtained using the source signature approach. The second was a check on the quality of the source and receiver signals, which involved viewing the resulting source signature and sampling of receiver waveforms from all 275 source locations and manually removing source impacts that were of very poor quality. Source records were removed if they were contaminated with significant high-amplitude noise, such as from trucks driving down the nearby highway or anthropogenic activity in the storage yard to the south of the array. While quite tedious and time consuming, this process of removing noisy source impacts was determined to be worthwhile because it resulted in substantially cleaner waveforms. As a result, while three vibroseis and five sledgehammer shots were performed at most source locations, only those deemed of good quality were included in the stacking process. Example waveforms will be presented in the next section of the paper on potential use cases of the dataset.

The last step of the receiver data post-acquisition processing was to convert the data to engineering units and provide the results in an easy-to-reuse data format. The process of converting data measured with geophones to engineering units is not as straightforward as those described above for the vibroseis and sledgehammer sources, as the geophone's amplitude and phase response are frequency dependent. To correct for the instrument's response, its frequency-dependent amplitude and phase responses must be removed from the recorded waveforms. This is especially important for frequencies near and below the sensor's resonant frequency, which for the sensors used in this study is 5 Hz. To remove this frequency-dependent effect, IRIS PASSCAL provides the frequency response of each sensor from the manufacturer. Under ideal circumstances, the measured frequency response of each component of each sensor would be used rather than the standard manufacturer-provided response. However, this was not possible for this study, and so we rely on the manufacturer-provided response. The response of each component of these nodal stations can be described with a sensitivity of 4.12E+09 m/s per count, gain of 0.9998, poles at [-21.99 + 22.43j, -21.99 – 22.43j ], and zeros at [0+0j, 0-0j], where $j$ is the imaginary number (i.e., $\sqrt{-1}$ ). Note the instrument response information has also been included with the published dataset in both StationXML and RESP formats. The conversion to physical units was performed using the $simulate\_seismometer$ function from the \emph{obspy} Python package (The ObsPy Development Team, \citeyear{the_obspy_development_team_obspy_2022}). In addition to the information listed previously, a cosine taper over 5\% of the sensor signal (2.5\% off either end) and a high pass filter at 3 Hz with a cosine taper to zero at 2 Hz was applied to mitigate frequency-domain artifacts during the correction process. The results of this process are signals that as closely as possible measure the ground's true particle velocity in engineering units of m/s. Note that due to the somewhat subjective nature of this process, including the size of the cosine taper and the use of a high pass filter and its corresponding frequencies, we have provided both the raw data in units of counts (i.e., without any instrument correction performed) and the data in units of m/s (i.e., after the instrument correction as described above has been performed). We hope that by providing the data in both formats, we enable users who agree with the choices made above to easily reuse the data, but also accommodate those who wish to use a different instrument correction procedure. The uncorrected and corrected data have been saved in CSV format. The choice to use the CSV format rather than a binary based format, such as was used for the passive data, was due to the plethora of potential formats for active-source data used in geophysics (e.g., SEG-Y rev2, SEG-Y rev3, SEG-2, SU, etc.) and the fact that there is no clear industry standard for active-source data unlike the de-facto standard for passive data that is miniSEED. The authors felt that the selection of any one active-source format would likely make the data prohibitively difficult to use by those whose software uses another format by default, as open-source readers and writers for all the possible formats are not readily available. For those looking for a versatile binary format, the authors recommend using \emph{obspy} (The ObsPy Development Team, \citeyear{the_obspy_development_team_obspy_2022}) to write the data to the Seismic Unix (SU) format. As a result, for each source impact, there are four associated CSV files (one for the source signature and three for the stations (one per component)). In all files the first column provides a time vector followed by one column for each associated receiver waveform. The header of each column specifies the type of data recorded and in parenthesis the units of that data. The entire active-source dataset, despite being stored in a text based format and including the data in both counts and engineering units, is only 25 GB and includes approximately 9k files. The next and final section of this paper details potential use cases of the passive-wavefield and active-source data.

\section*{Potential Dataset Use Cases}

\subsection*{Three-Dimensional Subsurface Imaging via FWI and ANT}

\begin{figure}[!]
    \centering
	\includegraphics[width=1.0\textwidth]{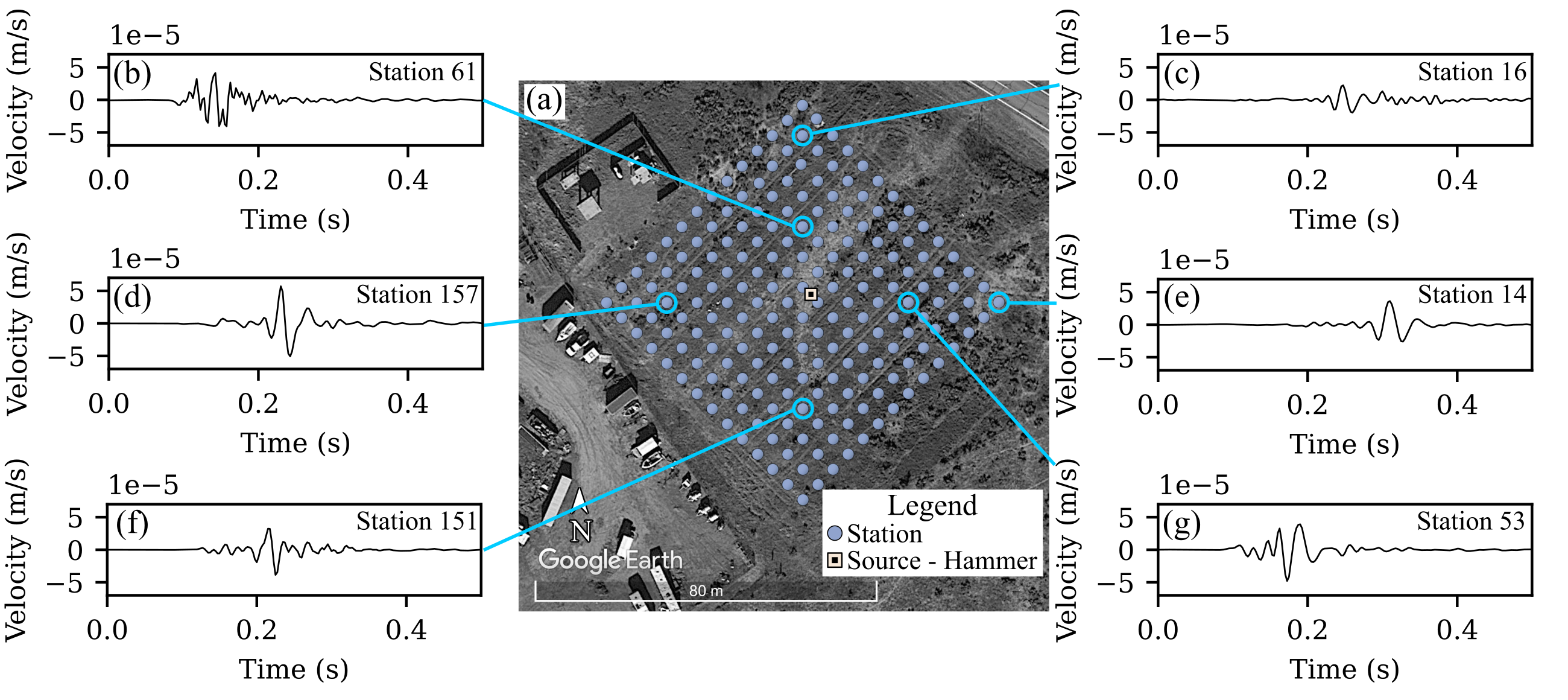}
	\caption{Examples of vertical-component waveforms recorded from a single, active-source hammer impact at location 78. Panel (a) shows the location of the example hammer source relative to six example recording locations. Panels (b) – (g) present the stacked vertical component in engineering units from the six stations (i.e., 61, 16, 157, 14, 151, and 53, respectively).}
	\label{fig:8}
\end{figure}

\begin{figure}[!]
    \centering
	\includegraphics[width=0.8\textwidth]{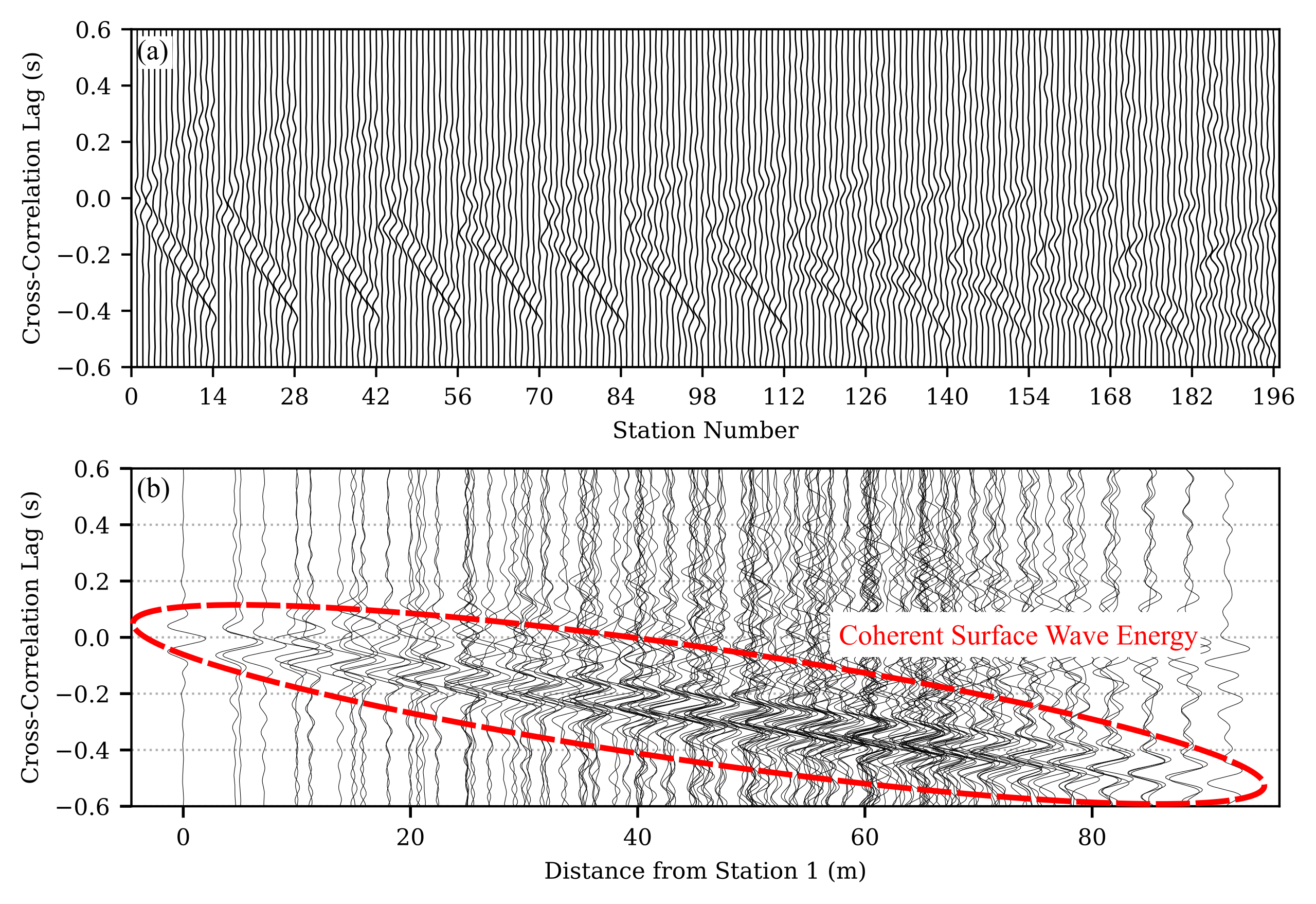}
	\caption{Average cross-correlation between the vertical components of all 196 stations and station 1 produced from four hours of ambient noise using a 1-second long time window. Results are presented by station number in panel (a) and by distance from the station 1 (i.e., the reference station) in panel (b).}
	\label{fig:9}
\end{figure}

To illustrate the use of the presented dataset for active-source 3D imaging experiments, such as FWI, Figure \ref{fig:8}a shows a plan view of a single hammer source location (i.e., 78; refer to Figure \ref{fig:5}) relative to a selection of six stations (i.e., 61, 16, 157, 14, 151, and 53; refer to Figure \ref{fig:3}). Figures \ref{fig:8}b to \ref{fig:8}g show the stacked vertical components of the six selected stations in engineering units, respectively. The recordings show that even a relatively weak sledgehammer source was able to produce clear waveforms with apparently good SNR propagated across the spatial extents of the array. To illustrate the value of the passive-wavefield measurements for 3D imaging experiments, such as ambient noise tomography (ANT), Figure \ref{fig:9} presents the cross-correlation waveforms for four hours of ambient noise between all stations and station one. More specifically, four hours of ambient noise (UTC: $2019\_10\_09T10$ to $2019\_10\_09T13$) recorded by each station was filtered between 2 and 15 Hz, divided into 3600, one-second long time windows, and each window cross-correlated with the corresponding window recorded on station 1. The resulting 3600 sets of cross-correlation time series for each stations were averaged together to produce the average cross-correlation lags presented in Figure \ref{fig:9}a by station number and Figure \ref{fig:9}b by distance from station 1. As expected, the coherent surface wave energy is observed to be much strong at negative time lags, indicating wave energy primarily arriving from the northeast (i.e., in the direction of the nearby highway).

\subsection*{Horizontal-to-Vertical Spectral Ratio (HVSR)}

\begin{figure}[!]
    \centering
	\includegraphics[width=1.0\textwidth]{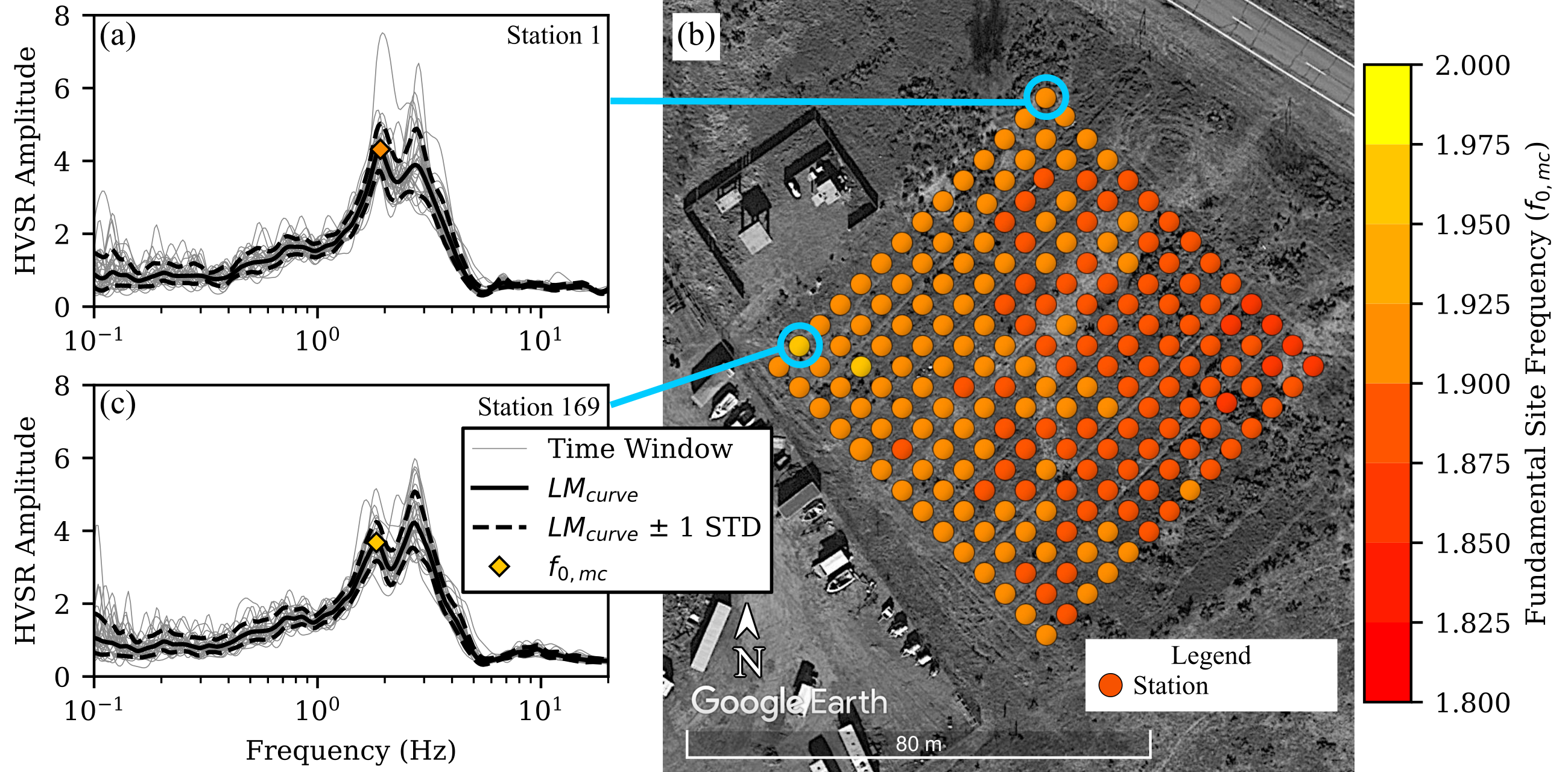}
	\caption{Example horizontal-to-vertical spectral ratio (HVSR) data calculated using one hour of ambient noise. Panels (a) and (c) show detailed plots of HVSR curves computed at stations 1 and 169, respectively, including the individual HVSR curves from each 120-second time window, the lognormal median curve ($LM_{curve}$) $\pm$ 1 lognormal standard deviation (STD), and the fundamental site frequency from the median curve ($f_{0,mc}$). Panel (b) shows the spatial distribution of $f_{0,mc}$ across all stations.}
	\label{fig:10}
\end{figure}

The passive-wavefield dataset can also be used to calculate horizontal-to-vertical spectral ratios (HVSRs). To illustrate, Figure \ref{fig:10}b shows the 196 stations color-mapped in terms of their fundamental natural frequency from the HVSR median curve ($f_{0,mc}$) using HVSR computed from one hour (UTC: $2019\_10\_09T10$) of ambient noise. The calculation was performed using the open-source Python package \emph{hvsrpy} \citep{vantassel_jpvantasselhvsrpy_2021} and involved dividing the one-hour-long recording into 120-second-long time windows, combining the horizontal components using the geometric mean, as recommended by Cox et al. (\citeyear{cox_statistical_2020}), smoothing with the filter proposed by Konno and Ohmachi (\citeyear{konno_ground-motion_1998}) with b=40, and calculating the lognormal median (LM) and ± one lognormal standard deviation (STD) curves. The peak of the lognormal median curve ($f_{0,mc}$) was used for simplicity instead of the more rigorous calculation of fundamental frequency from the peaks of individual time windows. From Figure \ref{fig:10}b, we observe some variability in $f_{0,mc}$  across the site, with values ranging from 1.83 to 1.95 Hz. Figures \ref{fig:10}a and \ref{fig:10}c show the calculated HVSR in greater detail for stations 1 and 169, respectively. In both examples, we observe two peaks in the HVSR; one at approximately 2 Hz and one at approximately 3 Hz. The authors note that these closely spaced peaks may be indicative of the two impedance contrasts between the AL/DG and the DG/GR that are known to exist at the GVDA site. However, we have also observed azimuthal variability in these peaks and they may correspond to valley longitudinal versus valley perpendicular directions. Thus, further investigations are required to more clearly understand the meanings of the HVSR curves.

\subsection*{Multichannel Analysis of Surface Waves (MASW)}

\begin{figure}[!]
    \centering
	\includegraphics[width=1.0\textwidth]{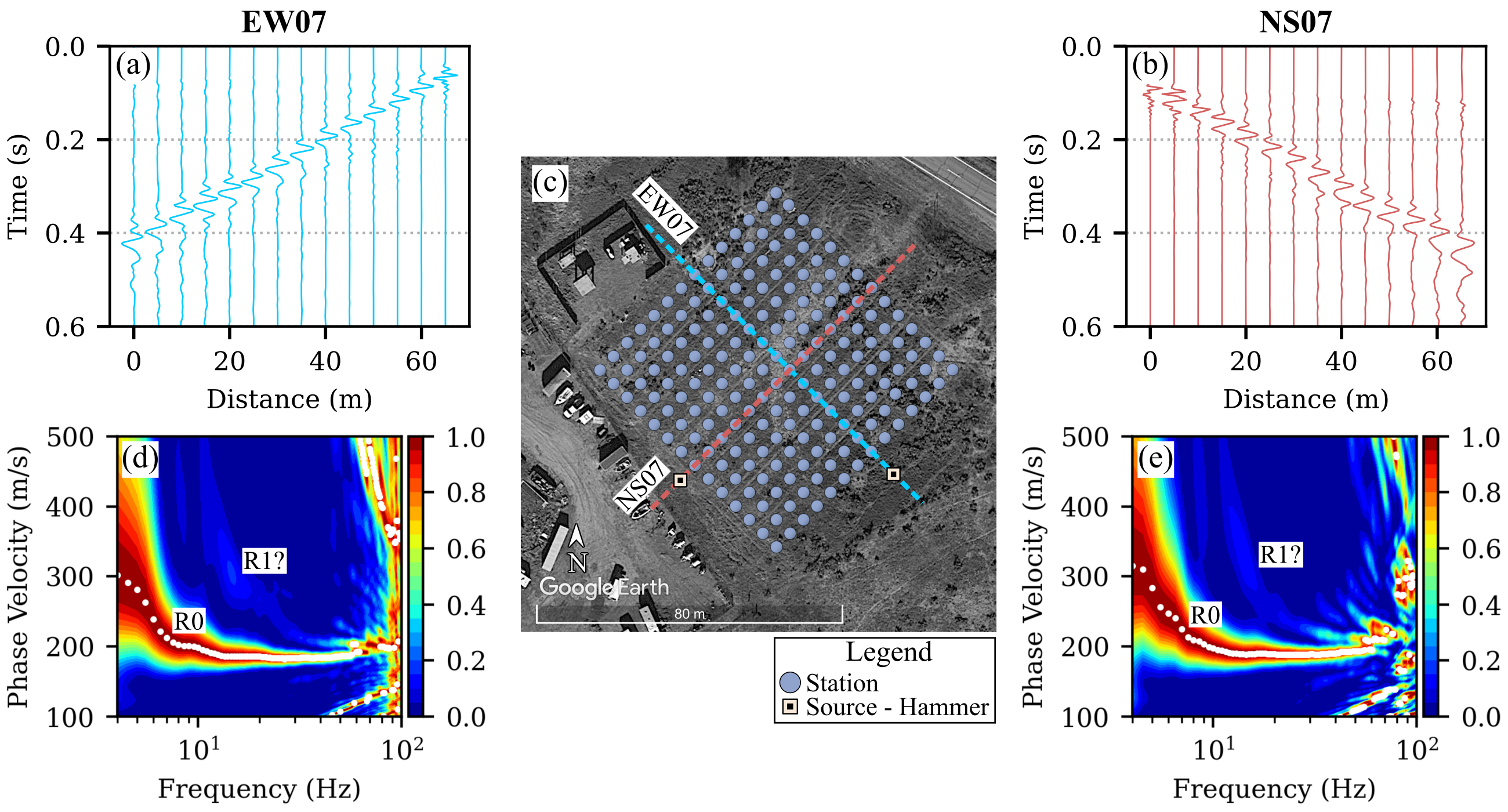}
	\caption{Example multichannel analysis of surface waves (MASW) data using two exterior hammer source locations. Panels (a) and (b) show the stacked vertical component waterfall plots associated with lines EW07 and NS07 for hammer source locations 134 and 160, respectively. Panel (c) shows a plan view of the GVDA site with the aforementioned linear arrays and shot locations indicated. Panels (d) and (e) show the dispersion images obtained from the waveforms in (a) and (b), respectively, after being transformed to the frequency-phase velocity domain using the frequency-domain beamformer with cylindrical-wave steering vector and square-root-distance weighting. The fundamental Rayleigh mode (R0) and potentially the first-higher Rayleigh mode (R1?) are labeled accordingly.}
	\label{fig:11}
\end{figure}

The hammer and vibroseis source locations positioned outside of the array (i.e., the exterior shots) can be used for multichannel analysis of surface waves (MASW) as a means to invert for shallow 1D shear wave velocity structure beneath each linear array of receivers. Figures \ref{fig:11}a and \ref{fig:11}b show time-domain stacked waveforms from two of the eight hammer source locations associated with lines EW07 and NS07, respectively. The location of each array line is illustrated Figure \ref{fig:11}c. Note the waveforms in Figures \ref{fig:11}a and \ref{fig:11}b are from source locations 134 and 160 (refer to Figure \ref{fig:5}), located 7.5 m away from the first geophone to the array's southeast and southwest, respectively. The surface wave dispersion images resulting from transforming the waveforms from Figures \ref{fig:11}a and \ref{fig:11}b are shown in Figures \ref{fig:11}d and \ref{fig:11}e. The wavefield transformation used is the frequency domain beamformer (FDBF) with cylindrical-wave steering vector and square-root-distance weighting, as developed by Zywicki and Rix (\citeyear{zywicki_mitigation_2005}) and implemented in the open-source Python package \emph{swprocess} \citep{vantassel_jpvantasselswprocess_2021}. The MASW data along both lines is broadband with clear, apparently fundamental mode Rayleigh (R0) dispersion data between 5 and 60 Hz. Due to the relatively large receiver spacing of 5 m, some spatial aliasing is observed at high frequencies ($>$s 50 Hz). The authors note that an apparently first-higher mode Rayleigh (R1) dispersion trend is also visible, although faint, in both dispersion images and has been denoted as (R1?) in the figures.

\subsection*{Microtremor Array Measurements (MAM)}

\begin{figure}[!]
    \centering
	\includegraphics[width=1.0\textwidth]{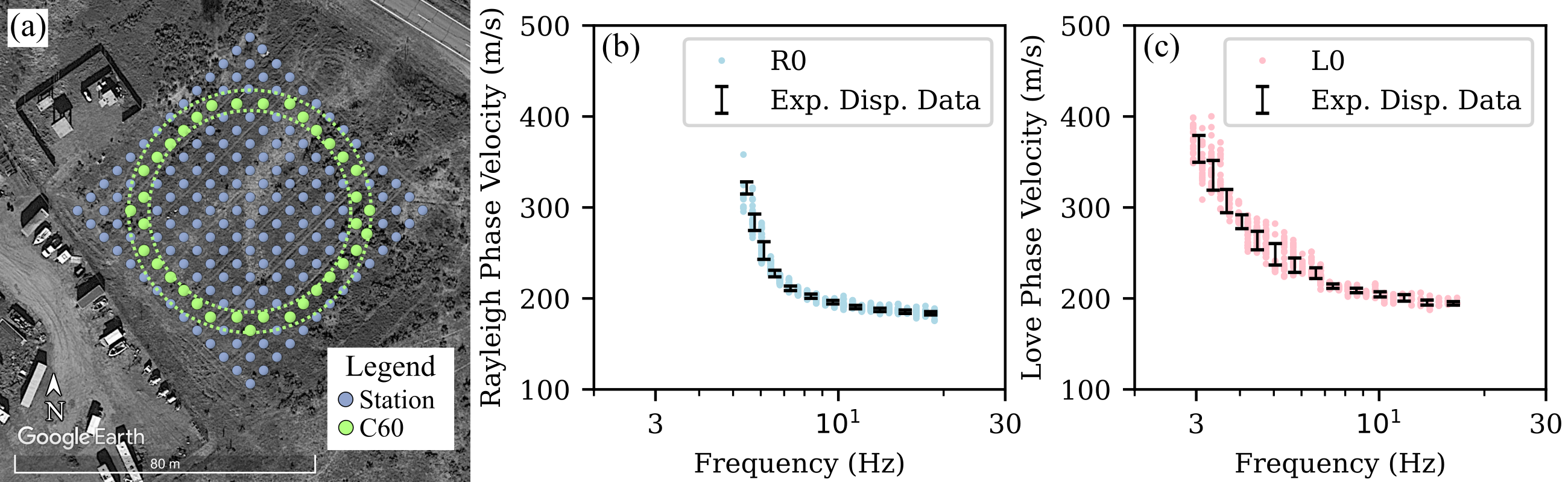}
	\caption{Example microtremor array measurements (MAM) made using one hour of passive-wavefield data from 36 of the 196 stations. The stations, which were selected to roughly correspond to a circular array with a diameter of 60 m (i.e., C60), are shown in panel (a). Panels (b) and (c) show the fundamental Rayleigh (R0) and fundamental Love (L0) wave dispersion data, respectively, extracted from the array noise measurements using the Rayleigh three-component beamformer (RTBF). Note that the dispersion data is shown after interactive trimming to remove outliers and alongside its associated mean and one standard deviation range.}
	\label{fig:12}
\end{figure}

The passive-wavefield recordings can also be used for microtremor array measurements (MAM) as a means to invert for deep 1D Vs structure beneath the entire array. While all 196 stations and all 28 hours of ambient noise data can be processed simultaneously, the computational demands in terms of memory and runtime exceed those pertinent for this preliminary demonstration of the data. Instead, a sampling of the stations and one hour (UTC: $2019\_10\_09T10$) of ambient noise data will be used. To examine lower frequency dispersion data than from the MASW shown previously, a 60 m-diameter circular array (C60) was constructed from 36 of the 196 stations. Stations for the C60 array were selected by finding all stations inside of two bounding circles of diameter 54 and 64 m. The bounding circles and the 36 selected stations are shown in Figure \ref{fig:12}a. The ambient noise data was processed using the Rayleigh three-component beamformer (RTBF) developed by Wathelet et al. (\citeyear{wathelet_rayleigh_2018}) and implemented in the open-source software \emph{geopsy} \citep{wathelet_geopsy_2020}. The RTBF processing used fixed-length, 30-second blocks combined into block sets of size 30. To accommodate the limited data used for this example (i.e., only one of 28 possible hours of noise), block sets were allowed to overlap up to 29 blocks. Up to five peaks were stored at each processing frequency. The results of the RTBF processing were post-processed using the open-source Python package \emph{swprocess} \citep{vantassel_jpvantasselswprocess_2021}. The fundamental model Rayleigh and Love wave dispersion data (i.e., R0 and L0) after interactive trimming to remove outliers and calculate statistics \citep{vantassel_swprocess_2022} are presented in Figures \ref{fig:12}b and \ref{fig:12}c, respectively. The R0 data is observed to be consistent with the MASW results for common frequencies (refer to Figures \ref{fig:11}d and \ref{fig:11}e). Furthermore, as expected from surface wave theory, the Rayleigh wave phase velocity data starts to cross below the Love wave data at high frequencies (approximately 13 Hz). The authors note that the Love wave data's extension to frequencies below that of the Rayleigh wave data indicates that Love wave data may help better constrain deeper Vs profiles developed at GVDA in the future.

\subsection*{Conclusions}

This paper presents a high quality, large-scale dataset for near-surface imaging acquired at the Garner Valley Downhole Array (GVDA) site. The dataset includes active-source and passive-wavefield recordings made using 196, 3-component nodal stations positioned on a 14 by 14 grid with a 5 m spacing. All stations were oriented to magnetic north and buried to ensure good coupling. The active-source data includes 66 vibroseis and 209 instrumented sledgehammer source locations. All active-source measurements include the recorded source signature and three-component station data in both counts and engineering units (i.e., kN) saved in an easy-to-reuse CSV format. Multiple source impacts were recorded at each source location and prepared for stacking but stored as separate CSV files (i.e., in their unstacked form) to enable future reuse. The passive-wavefield data includes 28 hours of ambient noise recorded over two night-time deployments. The passive-wavefield data is saved in one-hour time blocks for each station (all three components saved in a single file) in the easy-to-reuse miniSEED format. The dataset is shown to be useful for active-source and passive-wavefield 3D imaging as well as other subsurface characterization techniques, which include horizontal-to-vertical spectral ratios, multichannel analysis of surface waves, and microtremor array measurements. The authors hope that making this unique dataset available will help accelerate work in 3D subsurface imaging for near-surface engineering applications.

\subsection*{Acknowledgements}

The authors would like to acknowledge Alan Yong, Robert Kent, Dr. Mohamad Hallal, and Dr. Mauro Aimar for assisting in the field data acquisition. The authors would also like to thank Dr. Holly Rotman for answering the authors' questions regarding IRIS PASSCAL data and metadata conventions. This work was supported by the U.S. National Science Foundation grant CMMI-1931162. However, any opinions, findings, and conclusions or recommendations expressed in this material are those of the authors and do not necessarily reflect the views of the National Science Foundation. The figures in this paper were created using \emph{matplotlib} v3.5.2 and \emph{Inkscape} 1.1.2.

\subsection*{Conflict of Interest}

The authors declare no competing interests.

\bibliographystyle{plainnat}
\bibliography{gvda_data}

\begin{thebibliography}{31}
\providecommand{\natexlab}[1]{#1}
\providecommand{\url}[1]{\texttt{#1}}
\expandafter\ifx\csname urlstyle\endcsname\relax
  \providecommand{\doi}[1]{doi: #1}\else
  \providecommand{\doi}{doi: \begingroup \urlstyle{rm}\Url}\fi

\bibitem[Afshari et~al.(2019)Afshari, Stewart, and
  Steidl]{afshari_california_2019}
Kioumars Afshari, Jonathan~P. Stewart, and Jamison~H. Steidl.
\newblock California {Ground} {Motion} {Vertical} {Array} {Database}.
\newblock \emph{Earthquake Spectra}, 35\penalty0 (4):\penalty0 2003--2015,
  November 2019.
\newblock ISSN 8755-2930, 1944-8201.
\newblock \doi{10.1193/070218EQS170DP}.

\bibitem[Archuleta et~al.(1992)Archuleta, Seale, Sangas, Baker, and
  Swain]{archuleta_garner_1992}
Ralph~J. Archuleta, Sandra~H. Seale, Peter~V. Sangas, Lawrence~M. Baker, and
  Scott~T. Swain.
\newblock Garner {Valley} {Downhole} {Array} of {Accelerometers}:
  {Instrumentation} and {Preliminary} {Data} {Analysis}.
\newblock \emph{Bulletin of the Seismological Society of America}, 82\penalty0
  (4):\penalty0 1592--1621, August 1992.

\bibitem[Bielak et~al.(2012)Bielak, Fathi, Jeong, Kallivokas, Kucukcoban, Menq,
  and Stokoe]{bielak_t-rex_2012}
Jacobo Bielak, Arash Fathi, Chanseok Jeong, Loukas Kallivokas, Sezgin
  Kucukcoban, Farn-Yuh Menq, and Kenneth~H. Stokoe.
\newblock T-{Rex} shaking at {Garner} {Valley} - toward three-dimensional
  full-waveform inversion.
\newblock \emph{DesignSafe-CI [publisher]}, 2012.
\newblock \doi{10.4231/D3BK16P79}.

\bibitem[Bonilla et~al.(2002)Bonilla, Steidl, Gariel, and
  Archuleta]{bonilla_borehole_2002}
L.~F. Bonilla, J.~H. Steidl, J.~Gariel, and R.~J. Archuleta.
\newblock Borehole {Response} {Studies} at the {Garner} {Valley} {Downhole}
  {Array}, {Southern} {California}.
\newblock \emph{Bulletin of the Seismological Society of America}, 92\penalty0
  (8):\penalty0 3165--3179, December 2002.
\newblock ISSN 0037-1106.
\newblock \doi{10.1785/0120010235}.

\bibitem[Butzer et~al.(2013)Butzer, Kurzmann, and Bohlen]{butzer_3d_2013}
S.~Butzer, A.~Kurzmann, and T.~Bohlen.
\newblock {3D} elastic full‐waveform inversion of small‐scale
  heterogeneities in transmission geometry.
\newblock \emph{Geophysical Prospecting}, 61\penalty0 (6 - Challenges of
  Seismic Imaging and Inversion Devoted to Goldin):\penalty0 1238--1251, 2013.
\newblock ISSN 1365-2478.
\newblock \doi{https://doi.org/10.1111/1365-2478.12065}.
\newblock Publisher: European Association of Geoscientists \&amp; Engineers
  Type: Journal Article.

\bibitem[Cox et~al.(2020)Cox, Cheng, Vantassel, and
  Manuel]{cox_statistical_2020}
Brady~R Cox, Tianjian Cheng, Joseph~P Vantassel, and Lance Manuel.
\newblock A statistical representation and frequency-domain window-rejection
  algorithm for single-station {HVSR} measurements.
\newblock \emph{Geophysical Journal International}, 221\penalty0 (3):\penalty0
  2170--2183, June 2020.
\newblock ISSN 0956-540X, 1365-246X.
\newblock \doi{10.1093/gji/ggaa119}.

\bibitem[Fathi et~al.(2016)Fathi, Poursartip, Stokoe~II, and
  Kallivokas]{fathi_three-dimensional_2016}
Arash Fathi, Babak Poursartip, Kenneth~H. Stokoe~II, and Loukas~F. Kallivokas.
\newblock Three-dimensional {P}- and {S}-wave velocity profiling of
  geotechnical sites using full-waveform inversion driven by field data.
\newblock \emph{Soil Dynamics and Earthquake Engineering}, 87:\penalty0 63--81,
  August 2016.
\newblock ISSN 02677261.
\newblock \doi{10.1016/j.soildyn.2016.04.010}.

\bibitem[Gibbs(1989)]{gibbs_near-surface_1989}
J.~F. Gibbs.
\newblock Near-surface {P}- and {S}-wave velocities from borehole measurements
  near {Lake} {Hemet}, {California}.
\newblock Open-{File} {Report} 89-630, U.S. Geologic Survey, 1989.
\newblock URL \url{https://doi.org/10.3133/ofr89630}.

\bibitem[Hill(1981)]{hill_geology_1981}
Robert~I Hill.
\newblock Geology of {Garner} {Valley} and vicinity.
\newblock 1981.
\newblock Publisher: South Coast Geological Society.

\bibitem[Konno and Ohmachi(1998)]{konno_ground-motion_1998}
Katsuaki Konno and Tatsuo Ohmachi.
\newblock Ground-{Motion} {Characteristics} {Estimated} from {Spectral} {Ratio}
  between {Horizontal} and {Vertical} {Components} of {Microtremor}.
\newblock 88\penalty0 (1):\penalty0 228--241, 1998.

\bibitem[Pecker and Mohammadioun(1991)]{pecker_downhole_1991}
A.~Pecker and B.~Mohammadioun.
\newblock Downhole {Instrumentation} for the {Evaluation} of {Non}-linear
  {Soil} {Response} on {Ground} {Surface} {Motion}.
\newblock K - {Seismic} {Response} {Analysis} and {Design}, Tokyo, Japan,
  August 1991. IASMiRT.

\bibitem[Rathje et~al.(2017)Rathje, Dawson, Padgett, Pinelli, Stanzione, Adair,
  Arduino, Brandenberg, Cockerill, Dey, Esteva, Haan, Hanlon, Kareem, Lowes,
  Mock, and Mosqueda]{rathje_designsafe_2017}
Ellen~M. Rathje, Clint Dawson, Jamie~E. Padgett, Jean-Paul Pinelli, Dan
  Stanzione, Ashley Adair, Pedro Arduino, Scott~J. Brandenberg, Tim Cockerill,
  Charlie Dey, Maria Esteva, Fred~L. Haan, Matthew Hanlon, Ahsan Kareem, Laura
  Lowes, Stephen Mock, and Gilberto Mosqueda.
\newblock {DesignSafe}: {New} {Cyberinfrastructure} for {Natural} {Hazards}
  {Engineering}.
\newblock \emph{Natural Hazards Review}, 18\penalty0 (3):\penalty0 06017001,
  August 2017.
\newblock ISSN 1527-6988, 1527-6996.
\newblock \doi{10.1061/(ASCE)NH.1527-6996.0000246}.

\bibitem[Steidl et~al.(1996)Steidl, Tumarkin, and Archuleta]{steidl_what_1996}
Jamison~H Steidl, Alexei~G Tumarkin, and Ralph~J Archuleta.
\newblock What {Is} a {Reference} {Site}?
\newblock page~16, 1996.

\bibitem[Steller(1996)]{steller_new_1996}
Robert Steller.
\newblock New {Borehole} {Geophysical} {Results} at {GVDA}.
\newblock Data {Report}, March 1996.

\bibitem[Stokoe et~al.(2004)Stokoe, Kurtulus, and Menq]{stokoe_sasw_2004}
Kenneth~H. Stokoe, Asli Kurtulus, and Farn-Yuh Menq.
\newblock {SASW} {Measurements} at the {NEES} {Garner} {Valley} {Test} {Site},
  {California}.
\newblock Data {Report}, The University of Texas at Austin, Austin, Texas,
  January 2004.

\bibitem[Stokoe et~al.(2020)Stokoe, Cox, Clayton, and
  Menq]{stokoe_nheriutexas_2020}
Kenneth~H. Stokoe, Brady~R. Cox, Patricia~M. Clayton, and Farnyuh Menq.
\newblock {NHERI}@{UTexas} {Experimental} {Facility} {With} {Large}-{Scale}
  {Mobile} {Shakers} for {Field} {Studies}.
\newblock \emph{Frontiers in Built Environment}, 6:\penalty0 575973, November
  2020.
\newblock ISSN 2297-3362.
\newblock \doi{10.3389/fbuil.2020.575973}.

\bibitem[Tao and Rathje(2019)]{tao_insights_2019}
Yumeng Tao and Ellen Rathje.
\newblock Insights into {Modeling} {Small}-{Strain} {Site} {Response} {Derived}
  from {Downhole} {Array} {Data}.
\newblock \emph{Journal of Geotechnical and Geoenvironmental Engineering},
  145\penalty0 (7):\penalty0 04019023, July 2019.
\newblock ISSN 1090-0241, 1943-5606.
\newblock \doi{10.1061/(ASCE)GT.1943-5606.0002048}.

\bibitem[Teague et~al.(2018)Teague, Cox, and Rathje]{teague_measured_2018}
David~P. Teague, Brady~R. Cox, and Ellen~M. Rathje.
\newblock Measured vs. predicted site response at the {Garner} {Valley}
  {Downhole} {Array} considering shear wave velocity uncertainty from borehole
  and surface wave methods.
\newblock \emph{Soil Dynamics and Earthquake Engineering}, 113:\penalty0
  339--355, October 2018.
\newblock ISSN 02677261.
\newblock \doi{10.1016/j.soildyn.2018.05.031}.

\bibitem[Team(2022)]{the_obspy_development_team_obspy_2022}
The ObsPy~Development Team.
\newblock {ObsPy} 1.3.0, March 2022.

\bibitem[Tran et~al.(2019)Tran, Mirzanejad, McVay, and Horhota]{tran_3-d_2019}
Khiem~T Tran, Majid Mirzanejad, Michael McVay, and David Horhota.
\newblock 3-{D} time-domain {Gauss}–{Newton} full waveform inversion for
  near-surface site characterization.
\newblock \emph{Geophysical Journal International}, 217\penalty0 (1):\penalty0
  206--218, April 2019.
\newblock ISSN 0956-540X, 1365-246X.
\newblock \doi{10.1093/gji/ggz020}.

\bibitem[Vantassel and Cox(2019)]{vantassel_multi-reference-depth_2019}
J~P Vantassel and B~R Cox.
\newblock Multi-reference-depth site response at the {Garner} {Valley}
  {Downhole} {Array}.
\newblock In \emph{Proceedings of the {VII} {ICEGE}}, page~8, Rome, Italy, June
  2019. CRC Press / Balkema, Taylor \& Francis Group.

\bibitem[Vantassel(2021{\natexlab{a}})]{vantassel_jpvantasselhvsrpy_2021}
Joseph Vantassel.
\newblock jpvantassel/hvsrpy: v1.0.0, October 2021{\natexlab{a}}.

\bibitem[Vantassel(2021{\natexlab{b}})]{vantassel_jpvantasselswprocess_2021}
Joseph Vantassel.
\newblock jpvantassel/swprocess: v0.1.0b0, March 2021{\natexlab{b}}.

\bibitem[Vantassel and Cox(2022)]{vantassel_swprocess_2022}
Joseph~P. Vantassel and Brady~R. Cox.
\newblock {SWprocess}: a workflow for developing robust estimates of surface
  wave dispersion uncertainty.
\newblock \emph{Journal of Seismology}, April 2022.
\newblock ISSN 1383-4649, 1573-157X.
\newblock \doi{10.1007/s10950-021-10035-y}.

\bibitem[Vantassel et~al.(2023)Vantassel, Cox, and
  Crocker]{vantassel_active-source_2023}
J.P. Vantassel, B.R. Cox, and J.~A. Crocker.
\newblock Active-{Source} and {Passive}-{Wavefield} 3-{Component} {Nodal}
  {Station} {Measurements} at the {Garner} {Valley} {Downhole} {Array}.
\newblock \emph{DesignSafe-CI [publisher]}, 2023.
\newblock \doi{10.17603/ds2-ngdh-af80}.
\newblock Type: dataset.

\bibitem[Wang et~al.(2019)Wang, Huang, Liu, Li, Yong, and Yang]{wang_3d_2019}
Zi-Ying Wang, Jian-Ping Huang, Ding-Jin Liu, Zhen-Chun Li, Peng Yong, and
  Zhen-Jie Yang.
\newblock {3D} variable-grid full-waveform inversion on {GPU}.
\newblock \emph{Petroleum Science}, 16\penalty0 (5):\penalty0 1001--1014,
  October 2019.
\newblock ISSN 1672-5107, 1995-8226.
\newblock \doi{10.1007/s12182-019-00368-2}.

\bibitem[Wathelet et~al.(2018)Wathelet, Guillier, Roux, Cornou, and
  Ohrnberger]{wathelet_rayleigh_2018}
M~Wathelet, B~Guillier, P~Roux, C~Cornou, and M~Ohrnberger.
\newblock Rayleigh wave three-component beamforming: signed ellipticity
  assessment from high-resolution frequency-wavenumber processing of ambient
  vibration arrays.
\newblock \emph{Geophysical Journal International}, 215\penalty0 (1):\penalty0
  507--523, October 2018.
\newblock ISSN 0956-540X, 1365-246X.
\newblock \doi{10.1093/gji/ggy286}.

\bibitem[Wathelet et~al.(2020)Wathelet, Chatelain, Cornou, Giulio, Guillier,
  Ohrnberger, and Savvaidis]{wathelet_geopsy_2020}
Marc Wathelet, Jean-Luc Chatelain, Cécile Cornou, Giuseppe~Di Giulio, Bertrand
  Guillier, Matthias Ohrnberger, and Alexandros Savvaidis.
\newblock Geopsy: {A} {User}-{Friendly} {Open}-{Source} {Tool} {Set} for
  {Ambient} {Vibration} {Processing}.
\newblock \emph{Seismological Research Letters}, April 2020.
\newblock ISSN 0895-0695, 1938-2057.
\newblock \doi{10.1785/0220190360}.

\bibitem[Yong(2019)]{yong_data_2019}
Alan Yong.
\newblock Data {Acquisition} for 2020 {Blind} {Trial} of {Surface}-based {Site}
  {Characterization} {Methods}.
\newblock \emph{International Federation of Digital Seismograph Networks
  [publisher]}, 2019.
\newblock \doi{10.7914/SN/6H_2019}.

\bibitem[Youd et~al.(2004)Youd, Bartholomew, and
  Proctor]{youd_geotechnical_2004}
T.~Leslie Youd, Hannah A.~J. Bartholomew, and Jacob~S. Proctor.
\newblock Geotechnical {Logs} and {Data} from {Permanently} {Instrumented}
  {Field} {Sites}: {Garner} {Valley} {Downhole} {Array} ({GVDA}) and {Wildlife}
  {Liquefaction} {Array} ({WLA}).
\newblock Data {Report}, Brigham Young University, Provo, UT, December 2004.

\bibitem[Zywicki and Rix(2005)]{zywicki_mitigation_2005}
Daren~J. Zywicki and Glenn~J. Rix.
\newblock Mitigation of {Near}-{Field} {Effects} for {Seismic} {Surface} {Wave}
  {Velocity} {Estimation} with {Cylindrical} {Beamformers}.
\newblock \emph{Journal of Geotechnical and Geoenvironmental Engineering},
  131\penalty0 (8):\penalty0 970--977, August 2005.
\newblock ISSN 1090-0241, 1943-5606.
\newblock \doi{10.1061/(ASCE)1090-0241(2005)131:8(970)}.

\end{thebibliography}

\end{document}